# Regard Renormalization in QED as Functor between Categories

Zhongzhu Liu

Department of Physics, Huazhong University of Science and Technology, Wuhan 430074, People's Republic of China

**Abstract** To unify the quantum electrodynamics (QED) under the first principle which brings the renormalization unartifically, we study Feynman diagrams in QED according to the set theory and the category theory. We add the restriction on the electromagnetic interaction that a particle can and only can interact with particles which never interact with onself. Thus, fermiors (lines) in tree diagrams belong to sets, but belong to proper classes when they are in five primitive divergent diagrams. Fermiors, photons, fermior mass and charge compose together categories in which the group product of the local U(1) group and the propr Lorentz group is the morphism. There is the functor projecting the category containing fermiors in tree diagrams into the category containing fermior in divergent diagrams. Because proper classes have not the measure, this functor avoids the restriction of the measure. It can be regarded as the renormalization in QED and all renormalization constants have not a fixed magnitude.

It is well known that infinite appears annoyingly in the perturbation expansion of QED [1]. To get accurate results, people adopted the renormalization project to correct the primary theory. The starting point of QED is the Lagrangian of electromagnetic systems. Although the renormalization is the indivisible part of QED, it seem to have own start, the assumption about bare and physical quantities [2]. So, the present QED has two reciprocally independent starts. We puzzle why the the Nature is built so complex. Hence, we imagine that QED has a uniform frame. The frame starts from the first principle and the renormalization is its unartifical result.

The initiatory Lagrangian of QED includes four physical quantities, the fermion field $\psi_0$, the electromagnetic field $A_0$, the fermion mass $m_0$ and the fermion charge $e_0$. After the renormalization they become the renormalized quantities $\psi_R$, $A_R$, $m_R$ and $e_R$. In current knowledge, $\psi_0$ and the like are bare quantities, and $\psi_R$ and the like are physical quantities. There are the renormalization relations between them as follows [1,2]

$$\begin{aligned} \psi_R &\to \psi_0 = Z_2^{1/2} \psi_R \\ A_{R\mu} &\to A_{0\mu} = Z_3^{1/2} A_{R\mu} \\ e_R &\to e_0 = Z_1 e_R \\ m_R &\to m_0 = m_R - \delta m \end{aligned} \quad (1)$$

where $Z_1, Z_2, Z_3, \delta m$ are corresponding renormalization constants. On the other hand, we would like to regard the renormalization as the mapping from the renormalized quantities into those primal. If the renormalization is an inevitable result in some theory, the theory must result in the mapping (1) automatically. Thus, the theory should firstly be able to determine a rule, according which physical quantities appearing in Lagrangian will emerge in two patterns (in the present renormalization, they are bare and physical patterns). And then it can build a mapping between physical quantities in two patterns. In processing the renormalization, renormalization constants undergo iterative algebraic operation. These operations require these constants are finite. However, these constants are infinity in the present account. Here a contradictory emerges. A acceptable management to overcome the contradictory is that these renormalization constants in the mapping have not a fixed magnitude. Moreover, some symmetry of the Lagrangian, the Lorentz transformation and the local U(1) symmetry, must be reserved under the mapping.

Present QED is theoretically complete that its basic suppositions provided almost all reasonable results. If someone tries added one new supposition to the supposition system of QED, he will face the danger that the new supposition possibly lead to excess conclusions acting against experimental facts. So, we may only made little correction on some supposition of QED to get a uniform description of QED. The Lagrangian of QED has two parts, the free Lagrangian of the fermion field and the electromagnetic field and the interaction Lagrangian among fields. The free Lagrangian describes the existence of both the fields that it results in motion equations of fields. So, the theory of QED has three basic suppositions, the existence of the fermion field and the electromagnetic field, the interaction between fields and the renormalization. These are three independent each other suppositions. Apparently, the only scene on which we can play tricks is the field interaction. In this manuscript, we will add a restriction on the interaction, and then discuss why such treatment educe the renormalization to be the inevitable result according to the set theory and the theory of categories [3,4].

There are mainly two types of Feynman diagrams in the perturbation theory of QED [1,2]. One type includes all tree diagrams, every of which gives finite result. Other are all divergent diagrams. Each divergent diagram include at least one of five primitive divergent diagrams.

The set theory called collections of objects the classes [3]. Classes are divided into sets and proper classes which have different logistic significance. The study in the manuscript got that fermions appearing in tree diagrams belong to sets. But, fermions appearing in five primitive divergent diagrams should belong to proper classes, but never to sets. Hence, fermions appearing in divergent diagrams will possess different meaning from that of fermions in tree diagrams. These difference bring consequentially the mapping between fermions. At the same time, due to the local gauge invariance and the Lorentz covariation of the Lagrangian, this mapping should be enlarged to include the electromagnetic field, the mass and the charge of spinor particles. The mapping can take the form resembleing Eq.1. Moreover, the magnitude of an object is determined by the measure of the object. Every measure is defined on sets, but not on any proper class [6]. Above mapping appears between sets and proper classes, so it will not be restricted by a measure. The mapping can appears in all Feynman diagrams and we can regarde it as the renormalization. The involved constants can be called renormalization constants and they have not fixed magnitudes.

The description line of the manuscript is listed as follows.

In Sec.II, we well suggest to restrict the "interaction" to appear only between different particles. According to the set theory [3], we confirm that fermions in tree diagrams belong to sets, but they in five primitive divergent diagrams belong to proper classes.

In Sec.III, according to the theory of categories [4], we compound categories from fermions, photons, the charge and the mass of fermions, and take group product of local U(1) group and the propr Lorentz group to be the morphism of categories [5].

Prescribe the functor from the category of fermions in primitive divergent diagrams into the category of fermions in tree diagrams to have the same form as the mapping (1).

According to the measure theory [6], show that the constants appearing in the functor have not a fixed magnitude, so the functor can be regarded as the renormalization in QED.

II. Collections of fermions in Feynman diagrams are either sets or proper classes

Terms in perturbation expansion of QED are expressed by Feynman diagrams [2]. Feynman diagrams with infinite values are constituted by five primitive divergent diagrams drawn in Fig.1 and Fig.2 [1,2]. The interpretions of these diagrams are common. Fig.1a expresses the fermion (antifermion) self-energy, and Fig.1b expresses a divergent vertex that a fermion interacts with two photons. Fig.2a expresses the vacuum polarization, Fig.2b expresses the closed loop of three fermion lines and Fig.2c expresses the closed loop of four fermion lines. We note that the Furry's theorem claims that the contributions of the summation of closed loops in Fig.2b is zero, and the Ward equality determines that the contributions of the closed loop in Fig.2b is also zero [7]. Whereas, these conclusions resulted from the symmetry of the electromagnetic Lagrange, but not from the set viewpoint, so we still discuss these diagrams according to the set theory.

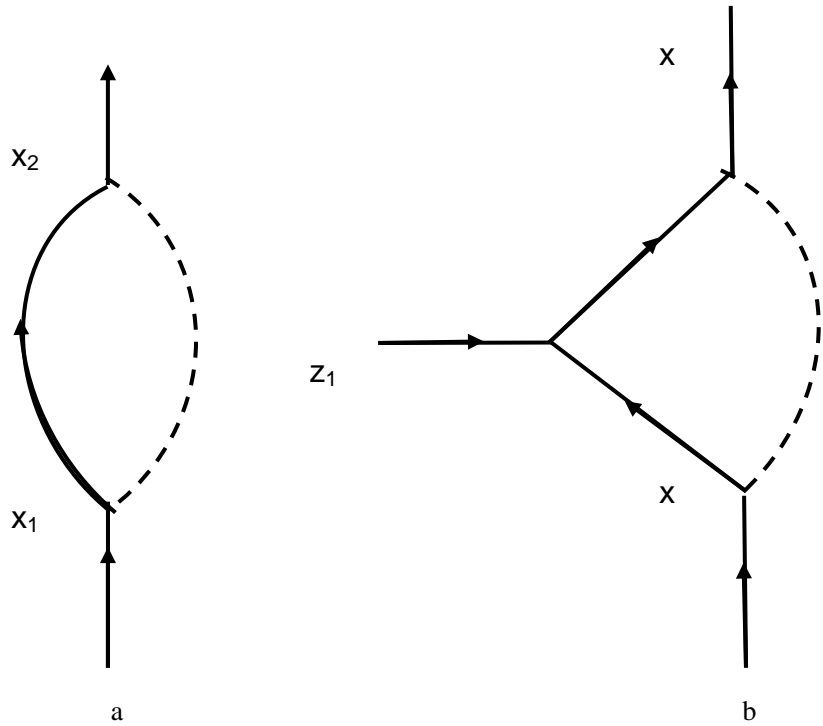

Fig.1 a. fermion (antiparticle) self-energy; b. vertex.

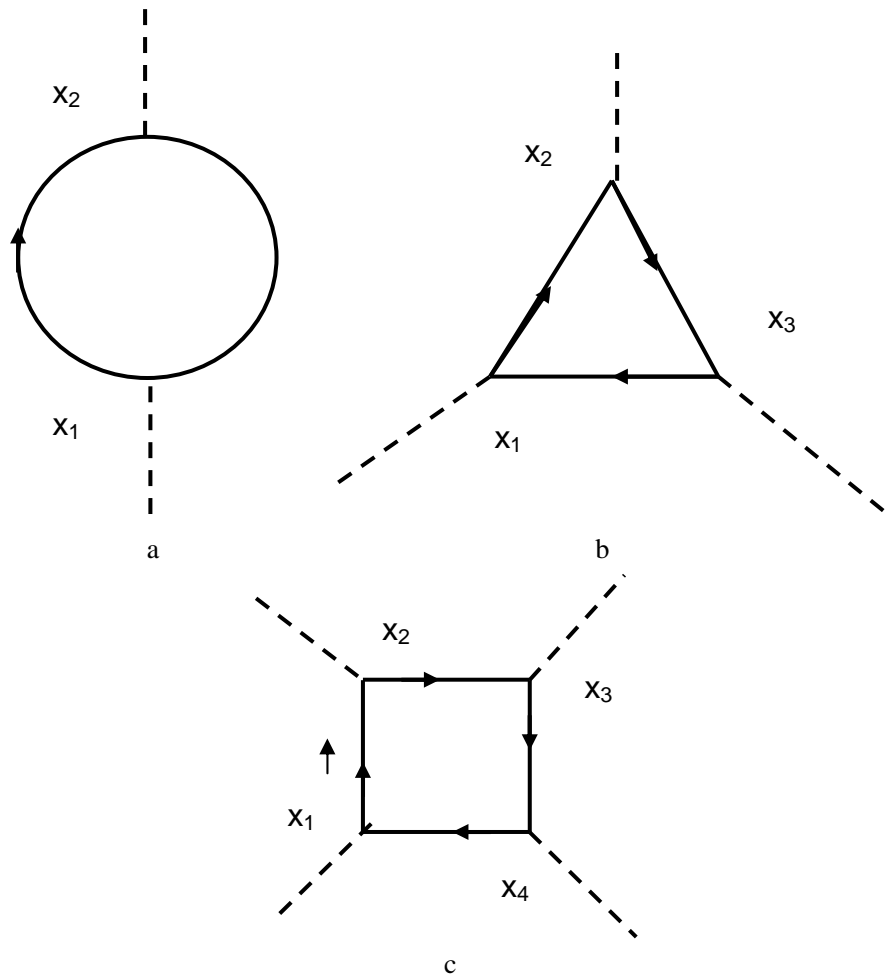

Fig.2 a. vacuum polarization; b. closed loop of three fermion lines; c. . closed loop of four fermion lines.

Feynman diagrams with finite values are tree diagrams which do not contain any of above primitive divergent diagrams.

The theory of QED is based on the theory of complex functions, the group theory and so on, which are built as well as on the set theory [8]. The property in the set theory of physical quantities is the base of all physical properties of the quantities. The difference in the set theory will determines these physical quantities are different. So, we study firstly the property in the set theory of physical quanlities in Feynman diagrams. In the following discussion, we will utilize the language of the predication logic applied in the set theory. This language appears in textbooks universally [9].

Firstly, we analyse the classical definition of the "interaction". The essence of this definition can be shown with the following example. We study a group of N electrons parking at fixed positions, which are the foundamental object in the Classical Electrodynamics [10]. Denoted the electron charge with $-e$ and the dielectric constant of the vacuum with $\varepsilon_0$, we can introduce the abbreviate $Q = \dfrac{e}{\sqrt{4\pi\varepsilon_0}}$.

Moreover, we record two electrons with $x$ and $y$ respectively, and the distance between electrons with r, then the Coluomb's force $f = Q^2 r^{-2}$ between $x$ and $y$ is written with predication sentences as follows:

Individual variables:   r, $r_1$, $r_2$, f;

Individual constants: $Q$, $G$;

Proposition variables:   $X$ is the collection of electrons;
  R(**,*,*) ** is the distance between * and *;
  N(*)： * is nozero;
  Q2(**, *)： ** is the square of *;
  M(**.*) ： ** is the inverse number of *;
  F(*) * is a force;
  P(**;*,*) ** is the product of * and *.

$\forall x \forall y (x \in X \wedge y \in X \rightarrow \forall r(R(r,x,y) \wedge N(r) \rightarrow (Q2(G,Q) \wedge Q2(r_1,r) \wedge M(r_2, r_1) \wedge \exists f(F(f) \wedge P(f;G,r_2)))))$.

(2)

The atomic formula in the above equation can be expressed with the symbol $A(x,y)$

$A(x,y) = (x \in X \wedge y \in X \rightarrow \forall r(R(r,x,y) \wedge N(r) \rightarrow (Q2(G,Q) \wedge Q2(r_1,r) \wedge M(r_2, r_1) \wedge \exists f(F(f) \wedge P(f;G,r_2)))))$.

Then the sentence (2) is rewritten as

$$\forall x \forall y A(x, y).  \quad (3)$$

The set theory describes the more abstract part of a notion. When our study focuses on the significance of the term "interaction" in the set theory, we can ignore the detail of the interaction. Then the Coluomb's force can be expressed as an ordered pair $\langle x, y \rangle$. In the Classical Electrodynamics, an accepted Coluomb's force appears only at the case $r \neq 0$, namely electrons $x$ and $y$ are not the same particle. Contrariwise, the ordered pair $\langle x, x \rangle$ corresponds with the self-action of the electron $x$. It give a infinite force. This is a puzzle in either concept or account. When face the puzzle, peaple usually use the additional prescription $r \neq 0$ to avoid it. However, a rigorous theoretical description of the term "interaction" should use a formula to remove the puzzle from the description. Therefore, we use the restrictive sentence for the definition of the interaction that an electron may and only may interact with the electron which never interact with onself. And then, we can rewrite the interaction as following sentence

$X = \{x | \text{interacting electrons}\}$

$\langle x, y \rangle = A(x, y)$

$$(\forall x)(x \in X \wedge (\forall y)(y \in X \rightarrow (\neg \langle y, y \rangle \leftrightarrow \langle x, y \rangle))). \quad (4)$$

where $\neg \langle y, y \rangle$ shows that electrons do no interact with onself. This sentence says that an electron x will interact with every electrons y if y does no interact with onself, The contradiction in (5) emerges for $x = y$ that the logical deduction shows

$$x = y \wedge (\neg \langle y, y \rangle \leftrightarrow \langle x, y \rangle)) \equiv$$

$$\{\neg \langle x, x \rangle \leftrightarrow \langle x, x \rangle\} \equiv$$
$$\{(\langle x, x \rangle \vee \langle x, x \rangle) \wedge (\neg \langle x, x \rangle \vee \neg \langle x, x \rangle)\} \equiv \{\langle x, x \rangle \wedge \neg \langle x, x \rangle\}. \quad (5)$$

The rusult $\{\langle x, x \rangle \wedge \neg \langle x, x \rangle\}$ says that the electron is able (expressed by the ordered pair $\langle x, x \rangle$) and synchronously unable (expressed by the negation of the ordered pair $\neg \langle x, x \rangle$) to interact with onself. In other words, an electron being unable to interact with onself will interact with onself, and an electron being able to interact with onself will be unable to interact with onself. The above sentence will not be dragged in any contradiction if it depicts the interaction only between different electrons, $x \neq y$. But, it will involve a paradox when it depicts whether an electron is able to interact with onself.

This is just the famous barber's paradox in the set theory [11] "There is a town with a barber who shaves all the people (and only the people) who don't shave themselves". The paradox is written with prediction sentences to be

$$(\exists x)\,(barber(x) \wedge (\forall y)(\neg shaves(y, y) \leftrightarrow shaves(x, y))). \qquad (6)$$

It can be recorded with the rigorous prediction sentence as follows

$Y = \{y | \text{peoples in the town}\}$

$$(\exists x)\,(x \in Y \wedge barber(x) \wedge (\forall y)(y \in Y \rightarrow (\neg shaves(y, y) \leftrightarrow shaves(x, y)))).$$

We can either express the formula $shaves(x, y)$ as an ordered pair

$$\langle x, y \rangle = shaves(x, y).$$

Then the sentence discribing the paradox is rewritten as follows

$Y = \{y | \text{peoples in the town}\}$

$\langle x, y \rangle = shaves(x, y)$

$z : barber$

$$.(\exists x)\,(x \in Y \wedge x = z \wedge (\forall y)(y \in Y \rightarrow (\neg \langle y, y \rangle \leftrightarrow \langle x, y \rangle))). \qquad (7)$$

At once, we see that Eq.(4) and Eq.(7) are the identical equations. So, a contradiction emerges equally well when $x = y$. The equation $x = y$ means the pair $\langle x, y \rangle$ becomes the pair $\langle x, x \rangle$. We can say also that the contradiction emerges along with the pair $\langle x, x \rangle$.

The barber's paradox is an embodiment of the Russell's paradox. In the set theory, the Russell's paradox is expressed as that a set $X$ is one own member

$$X \in X. \qquad (8)$$

Because Eq.(7) is same with Eq.(4), people will face the same problem about the self-action when it needs to definite a set of interacting electrons. The primal definetion of charge was introduced to describe the interaction between charged particles some distace apart. Afterwards, the definetion range of charge was extended to include the self-action of the charged particle itself [12]. However, this extension is doubtful that it brought an unreasonable infinite. So, the problem whether an electron may interaction with onself is same as the problem whether a barber may shave itself.

Feynman diagrams in QED are made up of fermion lines (antifermion lines) and photon lines as drawn in Fig.3a and Fig.3b. We describe Feynman diagrams with language of the set theory that regard an fermion line as an member $x$ of the fermion collection and regard a photon line as an ordered pair $\langle x, y \rangle$ that the photon is emitted by the fermion $x$ and is absorbed by the fermion $y$ (fermion lines). In the discussion according to the set theory, we can draw Feynman diagrams whether in the momentum picture or whether in the coordinate picture.

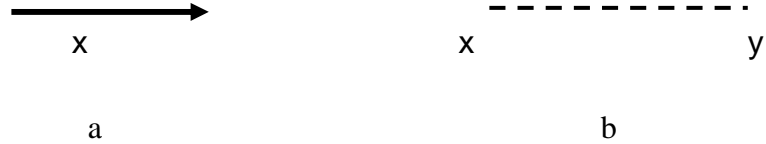

        a                           b

Fig.3 a. fermion line;   b. photon line.

For example, the Feynman diagram in the lowest order of the Møller scattering can be written as follows

$$(\exists x)\,(x \in X \wedge (\forall x')((x' \in X \wedge (\neg x = x')) \to \langle x, x' \rangle).$$

The self-energy figure of one fermion in Fig.1a shows that a fermion $x$ emits a photon while it arrives at the space-time point $x_1$, and then it arrives at the space-time point $x_2$ and absorbs the emitted photon. In the study about sets, we concern only what event happens, but not the idiographic process of the event. So, we can ignore the difference between classical and quantum motion. We now want to discribe the fermion in Fig.1a by a sentence of the set theory. This sentence is given as follows

$X = \{x | \text{interactive electrons}\}$

$$(\exists x)\,(x \in X \wedge \langle x, x \rangle). \tag{9}$$

The ordered pair $\langle x, x \rangle$ appears in the sentence. It means that the formula (9) will introduce a contradiction in a collection of interactive fermions according to the proof about Eq.(4). We can say that fermion collections appearing in the self-energy figure brought the Russell's paradox into the perturbation theory of QED. Or speaking, the collection of fermions in the self-energy figure possesses the property (8). The set theory shows that collections possessing the property (8) are proper classes, but not sets [3].

Similarly, the predication sentence describing the vertex figure in Fig.1b is

$$(\exists x)\,(x \in X \wedge \langle x, x \rangle \wedge \langle z_1, x \rangle). \tag{10}$$

where element $x$ is restricted and element $z_1$ is free which are fermions belonging to two respective collections. The orderin pair $\langle x, x \rangle$ expresses the inner photon line recording the self-action of a fermion. The orderin pair $\langle z_1, x \rangle$ expresses the external photon line that it has one restricted element and one free element. Comparing Eq. (9) and Eq.(10), we find that they have equally the ordered pair $\langle x, x \rangle$. In substances, they correspond with the barber's paradox in the set theory. So, fermions in the self-energy figure and the vertex figure all belong to proper calsses.

Now, we want to consider the closed firmion loops with three vertices in Fig.2b. Individual closed firmion loops is divergent and its S-matrix element involves the factor [1,2]

$$\gamma_\mu S_F(x_1 - x_2)\gamma_\lambda S_F(x_2 - x_3)\gamma_\sigma S_F(x_3 - x_1) \tag{11}$$

In accordance with the common physical interpretation, at least one of three propagators $S_F$ in the factor (11) is regarded as the inner line of an antiparticle. However, the propagator $S_F$ is unable to distiguish between fermions and antifermions. So, the figure can be regarded as a closed loop of three fermion inner lines as its usual appellation. Correspondingly, the closed loop has the relation

$$t_3 \geq t_1 \geq t_2 \geq t_3, \tag{12}$$

that is a fermion starting from the space-time point $(x, t_3)$, comes through moments $t_1$ and $t_2$ and recurs finally at the space-time point $(x, t_3)$. In Ref.13, we proven that an object must belong to a proper class if it moves along a closed timelike curve. In practice, the essence of our proof is independent of the trajectory of the object. So, the proof shows essentially that any object must belong to a proper class if it moves and recurs at the start obeying the relation (12). Since fermions in a closed loop of three fermion inner lines satisify the above condition, they belong to proper classes.

According to an analogous discussion, we can cognize that fermions have the relation $t_1 \leq t_2 \leq t_1$ when they are in the vacuum polarization graph Fig.2a and have the relation $t_1 \leq t_2 \leq t_3 \leq t_4 \leq t_1$ when they are in the closed loop of four fermions lines in Fig.2c. Fermions in such figures all recurred at the start, so they must belong to proper classes [13].

We can conclude according to the above discussion that fermions in primitive divergent diagrams all belong to proper classes if we restrict that the interaction only can appear between different particles.

On the other hand, fermions in tree diagrams can not be fazed by the Russell's paradox, they belong to still sets.

The set theory is the base of the mathematics and the physics [8], so fermions in different diagrams will be different mathematical objects. At this rate, the restriction on the interaction make a difference between fermions in tree diagrams and in primitive divergent diagrams that they belong to respectively sets and proper classes.

III. the renormalization is a functor between categories

In the above section, we proven that fermions in tree diagrams belong to sets, but fermions in primitive divergent diagrams belong to proper classes. Since sets and proper classes are different logistic objects, mathematics based on them must be different [3]. It means that fermions described by different logistic notions are different. So, fermions are divided into two group (two pattens) that they in the first group belong to sets, but they in the second group belong to proper classes. In this section, we will build the mapping between fermions in two groups.

The Lagrange of the QED is invariant under the local U(1) group and the proper Lorentz group $L_p$ [1]. The local U(1) symmetry associates a fermion fields with the electromegnetic field. Every element of the local U(1) group contains the charge $e$ to be parameter. Again, the fact that every fermion field is some spinor representation of the proper Lorentz group $L_p$ fastens the group and the fermion mass $m$. We can take group product $L_p \times U(1)$ to be the morphism among fermion fields. Hence, objects studied by us involve fermions, photons as well as the fermion mass and the charge.

Ferstly, we study fermion fields in tree diagrams which we denote by $\psi_t(x)$. Similarly, we denote electromagnetic fields in tree diagrams by $A_{t\mu}(x)$, the fermion mass by $m_t$ and the fermion charge by $e_t$. A Hilbert space may be regarded as a category [5]. Here, we build the category $\mathscr{M}$ with more objectes. Collecting the fermion field, the electromagnetic field, the charge and the fermion mass together to form the family $ob(\mathscr{M})$ of the category $\mathscr{M}$. A fermion field $\psi_t(x)$ is an object of the family $ob(\mathscr{M})$. Similarly, a electromagnetic field $A_{t\mu}(x)$ is also an object of the family. All together, every of the charge or the mass is also object of the family. The group product $L_p \times U(1)$ is the mapping between fermion fields as well as between electromagnetic fields. The charge and the mass are invariable under the group product. The local U(1) group and the proper Lorentz group have respective unit elements. Their product gives the unit element of the group product. Evidently, the group product obeys the associative law, bucause both of groups obeys the associative law and the product does not destroy the law. So, it obeys three axioms of the morphism [4,5]. Thereout, the family $ob(\mathscr{M})$ and the morphism form together the category $\mathcal{W}$. This definition of the category $\mathcal{W}$ is wider than that proposed for a quantum system in Fer.5.

We denote fermion fields in primitive divergent diagrams by $\psi_l(x)$, electromagnetic fields by $A_{l\mu}(x)$, the fermion mass by $m_l$ and the fermion charge by $e_l$. Equally, we can compose the family ob($\mathcal{C}$) from $\psi_l(x)$, $A_{l\mu}(x)$, $e_l$ and $m_l$. The family ob($\mathcal{C}$) and the group product $L_p \times U(1)$ form the other category $\mathcal{C}$. The fermion field a in primitive divergent diagrams belong to proper calsses, so the category $\mathcal{C}$ is a large category [4].

According to the theory of categories, a mapping between two different categories can be constructed [4]. This mapping between categories is called the functor. A functor is composed from the mapping between objects belong respectively to two categories and the mapping between morphisms. We can definite the functor $F$ from the category $\mathcal{C}$ into the category $\mathcal{W}$ as that the mappings between their objects are

$$\psi_l(x) \to \psi_t(x) = F\psi_l(x) = Z'^{1/2}_2 \psi_l(x)$$
$$A_{l\mu}(x) \to A_{t\mu}(x) = FA_{l\mu}(x) = Z'^{1/2}_3 A_{l\mu}(x)$$
$$e_l \to e_t = Fe_l = Z'_1 e_l$$
$$m_l \to m_t = Fm_l = m_l - \delta m' \tag{15}$$

where $Z'_1, Z'_2, Z'_3, \delta m'$ are quantities. Let $\alpha(e_l)$ be an element of the local U(1) group in the category $\mathcal{C}$. The functor maps it into the element $\alpha(e_t)$ of the local U(1) group in the category $\mathcal{W}$ that $\alpha(e_t) = F\alpha(e_l)$. Similarly, the functor maps a element $l(m_l)$ of $L_p$ the category $\mathcal{C}$ into $l(m_t)$. So, the functor $F$ between two morphisms is

$$F[\alpha(e_l) \cdot l(m_l)] = F\alpha(e_l) \cdot Fl(m_l) = \alpha(e_t) \cdot l(m_t). \tag{16}$$

Apparently, the unit element $I_t$ of the group product has

$$I_l \to I_t = FI_l \tag{17}$$

For two elements of the group product $\alpha_1(e_l) \cdot l_1(m_l)$ and $\alpha_2(e_l) \cdot l_2(m_l)$, there is the product

$$\alpha_3(e_l) \cdot l_3(m_l) = [\alpha_1(e_l) \cdot l_1(m_l)] \cdot [\alpha_2(e_l) \cdot l_2(m_l)]$$
$$= [\alpha_1(e_l) \cdot \alpha_2(e_l)] \cdot [l_1(m_l) \cdot l_2(m_l)]$$

According to the associative law of the group product. The product obeys

$$F[\alpha_3(e_l) \cdot l_3(m_l)] = F\{[\alpha_1(e_l) \cdot \alpha_2(e_l)] \cdot [l_1(m_l) \cdot l_2(m_l)]\} =$$
$$= F[\alpha_1(e_l) \cdot \alpha_2(e_l)] \cdot F[l_1(m_l) \cdot l_2(m_l)] =$$
$$= [\alpha_1(e_t) \cdot \alpha_2(e_t)] \cdot [l_1(m_t) \cdot l_2(m_t)] \quad . \quad (18)$$
$$= [\alpha_1(e_t) \cdot l_1(m_t)] \cdot [\alpha_2(e_t) \cdot l_2(m_t)] =$$
$$= F[\alpha_1(e_l) \cdot l_1(m_l)] \cdot F[\alpha_2(e_l) \cdot l_2(m_l)]$$

So, $F$ is a covariative functor [4].

The magnitude of an object or an individual is determined by a measure. According to the theory of the measure, a measure is built on sets [6]. Without doubt, mappings between sets are restricted by the measure of sets. The functor F defined by Eq.(15) and Eq.(16) is the mapping from the category $\mathcal{C}$ into the category $\mathcal{W}$. The category $\mathcal{C}$ is a large category containing proper calsses, on which we cannot build a measure [4,6]. So, there is no a measure restriction on the functor F. Correspondingly, all quantities $Z'_1, Z'_2, Z'_3, \delta m'$ in (15) have not a fixed magnitude.

Comparing Eq.1 and Eq.15, we can write the follwing equations

$$\begin{array}{lll} \psi_l = \psi_R & \psi_t = \psi_0 & Z_2 = Z'_2 \\ A_{l\mu} = A_{R\mu} & A_{t\mu} = A_{0\mu} & Z_3 = Z'_3 \\ e_l = e_R & e_t = e_0 & Z_1 = Z'_1 \\ m_l = m_R & m_t = m_0 & \delta m = \delta m' \end{array} \quad . \quad (19)$$

These equations show that the functor F is just a opration of the renormalization of QED and quantities $Z'_1, Z'_2, Z'_3, \delta m'$ are the renormalization constants.

IV. Discussion

In this work, we do not propose any new postulate and claim only a acceptable restriction on the electromagnetic interaction in QED. After taking the restriction, fermion fields in Feynman tree diagrams and divergent diagrams show different behaviors and belong respectively to sets or proper classes. With the morphism, the group product of the local U(1) group and the proper Lorentz group, categorys are composed of fermion fields, electromegnatic fields, the fermion charge and the fermion mass. Tree diagrams and divergent diagrams correspond with different categorys. A functor $F$ between two such categorys is definited. Because fermions in divergent diagrams belong to proper classes without a measure. Quantities $Z'_1, Z'_2, Z'_3, \delta m'$ appearing in the functor have not a fixed magnitude. We hereby regard the functor as operations of the renormalization of QED.

The configuration of our study is logical that it has not any new assumption and only imposes a restriction on one initial supposition of QED. This restriction results in two conclusions. One conclusion is that fermion fields in tree diagrams and divergent diagrams possess different meaning and a mapping between them exists. Second is that the concerned quantities by mapping have not a fixed magnitude. This restriction brought one new concept, proper classes. Fermion particles belonging proper classes appear in five primitive divergent. These divergent emerge only within the time determined by the uncertainty relation [1,2], while particles can not be observed immediately. So, this restriction does not bring any excess result against present experimental facts.

Reference

[1] N. N. Bogoliubov and D. V. Shirkov, *Introduction to the theory of Quantized Fields*, (Interscience, New York, 1959);

J. D. Bjorken and S. D. Drell, *Relativistic Quantum Mechanics*, and *Relativistic Quantum Fields*, (Mc. Graw-Hill, New York, 1964);

D. Lurie, *Particle and Fields*, (Interscience, New York, 1968);

J. Schwinger (ed) *Quantum Electrodynamics*, (Dover, New York, 1958).

[2] F. J. Dyson, *Phys. Rev.*, **75**, 486 (1949);

R. P. Feynman, *Phys. Rev.*, **76**, 749 (1949);

A. Salam, *Phys. Rev.*, **86**, 731 (1952).

[3] A. A. Fraenkel. and Bar-Hillel, *Foundations of Set Theory*, (North-Holland, Amsterdam, 1958);

R. B. Chanqui, *Axiomotic Set Theory, Impredicative Theories of Classes*, (North-Holland, Oxford, 1981);

A. P. Morse, *A Theory of Sets*, (Academic Press, New York, 1965);

R. Maddy, J. Symb. Log.*,* **48** 113 (1983).

[4] S. Mac Lane, *Categories for the Working Mathematician*, (2nd ed, Springer-Verlag New york Berlin Heidelberg, 1998 );

J. C. Qxtoby, *Mesure and Categories*, (2nd ed, Springer-Verlag New york Heidelberg Berlin, 1980 ).

[5] D. G. Holdsworth, *J. Philos. Logc.*, **6**, 441 (1977).

[6] D. J. Cohn, *Mesure Theory*, (Birkhauser, 1980).

[7] W. H. Furry, *Phys. Rev.*, **51**, 125 (1937);

J. C. Ward, *Phys. Rev.*, **78**, 1824 (1950).

[8] A. Robinson, *Non-Standard Analysis*, (North-Holland Publishing Company, 1974);

A. Mukherjes and K. Pothoven, *Real and Functional Analysis*, (Plenum Press, 1884);

N. Jacobson, *Lectures in Abstract Algebra*, V.1 (Van Nostrand 1951).

[9] J. Barwise: in *Handbook of Mathematical Logic*, edited by J. Barwise (North-Holland, Amsterdam, 1977), p. 6, A1.

[10] J. D. Jackson, *Classical Electrodynamics*, (John Wiley & Sons, Inc. 1999).

[11] W. V. Quine, Scientific American*,* April 1962, pp. 84–96.

[12] H. A. Lorentz, *The Theory Of Electrons*, (Cosimo Classics, New York, 2007).

[13] Zhongzhu Liu, (2011) arxiv:1106.4707 [math-ph].